\documentclass[twocolumn,aps,amsmath,amssymb]{revtex4}
\usepackage{mathrsfs}
\usepackage{epsfig,bm,dcolumn}
\usepackage{graphicx}
\usepackage{color}
\usepackage{amsmath}

\begin{document}

\title{Quantum normal-to-inhomogeneous superconductor phase transition in nearly two-dimensional metals}

\author{Heron \surname{Caldas}}
\email{hcaldas@ufsj.edu.br}
\affiliation{Departamento de Ci\^encias Naturais, Universidade Federal de S\~ ao Jo\~ao del Rei, \\
36301-160, S\~ao Jo\~ao del Rei, MG, Brazil}

\author{Mucio A. \surname{Continentino}}
\email{mucio@cbpf.br}
\affiliation{Centro Brasileiro de Pesquisas F\'{\i}sicas \\
Rua Dr. Xavier Sigaud, 150, Urca \\ 
22290-180, Rio de Janeiro, RJ, Brazil}
\date{\today}

\date{\today}

\begin{abstract}

In multi-band systems, electrons from different orbitals coexist at the Fermi surface. An attractive interaction among these quasi-particles gives rise to inter-band or hybrid pairs which eventually condense in a superconducting state. These quasi-particles
have a natural mismatch of their Fermi wave-vectors,  $\delta k_F$, which depends on the strength of the hybridization between 
their orbitals. The existence of this natural scale suggests the possibility of inhomogeneous superconducting ground states in these systems, even in the absence of an applied magnetic field. Furthermore, since hybridization $V$ depends on pressure, this provides an external parameter to control the wave-vectors mismatch at the Fermi surface. In this work, we study the phase diagram of a two-dimensional, two-band metal with inter-band pairing. We show that as the mismatch between the Fermi wave-vectors of the two hybrid bands is reduced, the system presents a normal-to-inhomogeneous superconductor quantum phase transition at a critical value of the hybridization $V_c=\Delta_0$. The superconducting ground state for $V<V_c$ is characterized by a wave-vector with magnitude $|\mathbf{q}_c|=q_c=2 \Delta_0/\bar{v}_f$. Here $\Delta_0$ is the superconducting gap in the homogeneous state and $\bar{v}_f$ the average Fermi velocity.   We discuss the nature of the quantum critical point (QCP) at $V_c$ and obtain the associated quantum critical exponents.

\end{abstract}


%\keywords{Suggested keywords}%Use showkeys class option if keyword display desired

\maketitle

\section{Introduction}
In a metal, an external magnetic field polarizes the quasi-particles giving rise to a mismatch of the Fermi wave-vectors of the different spin bands. In the presence of an attractive interaction between electrons with opposite spins, the ground state of the system is a homogeneous superconductor at least for sufficiently small mismatch $\delta k_F=k_F^{\uparrow}-k_F^{\downarrow}$. In the other limit of very large mismatches, we expect the system to be normal (non-superconducting) for physically reasonable attractive interactions.
An important and fruitful question in the study of superconductivity is the nature of the zero temperature phase diagram as the mismatch is reduced in the normal phase or, alternatively, how the BCS phase disappears as $\delta k_F$ increases.  A theoretical answer to this problem was given by Fulde and Ferrel \cite{Fulde64} and Larkin and Ovchinikov \cite{Larkin64}. These authors have shown the appearance of an intervening inhomogeneous superconducting phase between the homogeneous BCS and the normal phases as $\delta k_F$ increases. This phase generally known as a FFLO phase is  characterized by a wave-vector dependent order parameter which oscillates in space \cite{narduli}.  Different types of solutions associated with a single wave-vector can be obtained with different ground state energies \cite{Fulde64,Larkin64,narduli}. Here we consider the original Fulde-Ferrell ($FF$) or helicoidal solutions \cite{Fulde64},  such that, the order parameter is given by $\Delta= \Delta_0 e^{i \mathbf{q} \cdot \mathbf{x}}$. Experimentally, FFLO phases have been elusive and there is no conclusive evidence for their existence.

In cold atom systems, two fermionic species with different densities and attractive interactions  provides a controllable model system to search for FFLO phases \cite{Hulet,Drummond,KetterleReview}. The mismatch can be tuned by varying the number of atoms of the different species. These are clean systems in which the strength of the interaction between the different species can also be tuned. In spite of intensive search only for nearly one-dimensional systems evidence was found for FFLO type of correlations \cite{Hulet}.

In transition metals, rare-earths and actinides metals, electrons arising from different orbitals (s,p,d or f) coexist at  the Fermi surface \cite{pascoal}. These electrons hybridize and the new quasi-particles have  different Fermi wave-vectors which among other things depend on the strength of the hybridization \cite{igor,moreo,aline}. These systems and their inter-metallic compounds in the case the dominant attractive interaction is between quasi-particles  in different bands are natural candidates for a FFLO ground state, even in the absence of a magnetic field.  For a two-band metal, the mismatch $k_F^a-k_F^b$ is directly related to hybridization \cite{igor,aline}.
Since this can be controlled by pressure, we have an external  parameter which allows to probe the phase diagram and search for FFLO phases. Inter-band pairing may be particularly important in heavy fermion systems \cite{pascoal} where the dominant Kondo interaction between conduction electrons and the rather local $f$-electrons provides also a mechanism for inter-band attraction \cite{khomsky}. Notice that FFLO phases induced by pressure avoid all the complications associated with orbital effects due to an external magnetic field.

In this paper we study the phase diagram of a two-dimensional (2d), two-band system as a function of the mismatch between their Fermi wave-vectors controlled by the strength of their hybridization.  This is relevant for nearly two-dimensional materials with cylindrical Fermi surfaces. Some of our results are in common with  previous ones obtained for field induced FFLO phases in $2d$ materials at finite temperatures \cite{fflo2d,shimahara,Simmons,Rainer,combescot}. As in these studies we assume the existence of weak inter planar interactions so that the mean field BCS approach is justified at zero temperature \cite{shimahara}. However, these  works ignore the dynamical aspects inherent to the quantum critical phenomena. We give here a full treatment of the quantum phase transition from the normal-to-inhomogeneous  superconductor as the Fermi wave-vector mismatch is reduced from the normal phase. Since this $T=0$ transition is continuous or second order, it is associated with a quantum critical point (QCP) at a critical value $V_c$ of the hybridization. The instability of the normal state that we consider is that for a $FF$ superconducting state characterized by a single wave-vector $\mathbf{q}$. This is the first zero temperature instability as $V$ is reduced \cite{shimahara,nardulli}. Our approach yields the full thermodynamic behavior near the QCP allowing to identify the $FF$ or helicoidal superconducting phase from its precursors effects.

\section{Model Hamiltonian}

The Hamiltonian of the two-band system with hybridization and an inter-band interaction is given by \cite{igor},
\begin{eqnarray}
\label{eq-1}
{\cal H}&=&H-\sum_{k,\alpha}\mu_{\alpha}n_{\alpha}+H_{int}\\
\nonumber
H&=&\sum_{k \sigma} {\epsilon}^{a}_k a^{\dagger}_{k \sigma} a_{k \sigma}+\sum_{k \sigma} {\epsilon}^{b}_k b^{\dagger}_{k \sigma} b_{k \sigma} \nonumber  \\
&+&V \sum_{k \sigma}(a^{\dagger}_{k \sigma} b_{k  \sigma} +b^{\dagger}_{k  \sigma} a_{k\sigma}) \nonumber \\
H_{int}&=& g \sum_{k,k' \sigma} a^{\dagger}_{k' \sigma} b^{\dagger}_{-k'  -\sigma} 
b_{-k  \sigma} a_{k  -\sigma} \nonumber ,\,
\end{eqnarray}
where $a^{\dagger}_{k \sigma}$ ($b^{\dagger}_{k \sigma}$), $a_{k \sigma}$ ($b_{k \sigma}$) are creation and annihilation operators for the $a$ ($b$) particles  and ${\epsilon}^{\alpha}_k$ are their dispersion relations, defined by ${\epsilon}^{\alpha}_k=\xi_k^{\alpha}-\mu_{\alpha}$, where $\xi_k^{\alpha}=\frac{\hbar^2 k^2}{2m_{\alpha}}$ and $\mu_{\alpha}$ is the chemical potential of the (non-interacting) $\alpha$-particle ($\alpha=a,b$).  Since the inter-band interaction between particles $a$ and $b$ is attractive, we take $g <0$. In the case of a heavy fermion system, the $a$-band is a wide band of conduction electrons and $b$ a narrow band of $f$-electrons with effective mass $m_b \gg m_a$. Multi-band superconductivity was previously studied by Suhl et al. \cite{Suhl} and Moskalenko \cite{Moskalenko}. These authors differently from the present study considered intra-band interactions, with an inter-band term that transfers Cooper pairs between the bands. In our case superconductivity is associated with the existence of a finite anomalous correlation function, $<a_{i \sigma}^{\dagger }b_{i -\sigma}^{\dagger }>$. Since the conduction electrons are described by plane waves, which contain all the harmonics, this correlation function vanishes only in the normal phase.

Next we calculate the zero temperature response of the two-band system 
to a wave-vector and frequency dependent
\textit{fictitious field} that couples to the superconducting order parameter \cite{aline}
of interest, $<a_{i \sigma}^{\dagger }b_{i -\sigma}^{\dagger }>$. The  Hamiltonian associated with this coupling  is given by \cite{aline},
\begin{equation}
\mathcal{H}_{1}=-h_q \sum_{i}e^{i\mathbf{q}\cdot\mathbf{r}_{\mathbf{i}%
}}e^{i\omega _{0}t}(a_{i \sigma}^{\dagger }b_{i -\sigma}^{\dagger }+b_{i \sigma }a_{i -\sigma}),
\label{Hamiltoniano2}
\end{equation}
where the frequency $\omega $ has a small positive imaginary part
to guarantee the adiabatic switching on of the fictitious field. The linear response
of the system to this field, is obtained using
perturbation theory for the retarded and advanced Green's
functions \cite{aline}. We consider that in the two-band system  the hybridization is sufficiently large,  such that, in this initial condition the system is in the normal phase. In this case the
superconducting order parameter is zero in the absence of the
\textit{fictitious field} $h_q$.

\begin{figure}
\includegraphics[width=7.5cm]{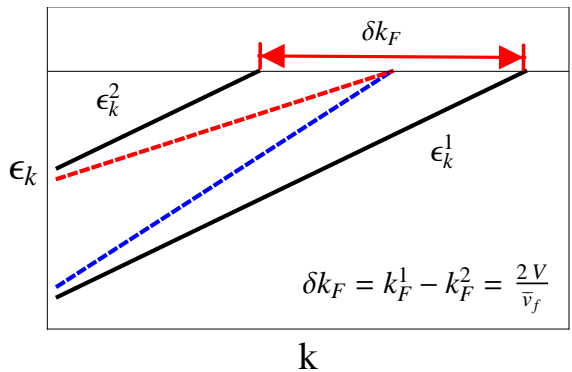}
\caption{(Color online) Simplified band-structure used in the calculation of the dynamic susceptibility. For the original bands unhybridized bands (dashed lines) we assume perfect nesting, such that, they cross the Fermi surface at the same wave-vector. The hybridized bands bands are shown as full lines and cross the Fermi surface at $k_F^1$ and $k_F^2$, with a mismatch $\delta k_F=k_F^1-k_F^2$.}
\label{fig1}
\end{figure}

The superconducting response to the time and $q$-dependent fictitious field, within the BCS decoupling
is obtained in the form \cite{aline},
\begin{equation}\label{gsinter}
\delta \Delta ^{ab}_q=\tilde{\chi}(q,\omega )h_q=\frac{\chi _{V}^{12}(q,\omega )}{1-g \chi
_{V}^{12}(q,\omega )}h_q,
\end{equation}%
where, $\delta \Delta^{ab}_q= \sum_k <a^{\dagger}_{k+q \sigma } b^{\dagger}_{-k-\sigma }>$, and the non-interacting  pair  susceptibility $\chi _{V}^{12}(q,\omega )$ is given by,
\begin{equation} \label{chi2d}
\chi _{V}^{12}(q,\omega )=\frac{1}{2\pi }\sum_{k}\frac{1-f(\epsilon
_{k-q}^{1})-f(\epsilon _{k+q}^{2})}{\epsilon _{k-q}^{1}+\epsilon
_{k+q}^{2}-\omega},
\end{equation}
where $1$ and $2$ refer to the new hybridized bands (see Fig.~\ref{fig1}), 
\[
\epsilon _{k}^{1,2}=\bar{v}_f (k-k_F)\mp V,
\]%
where $\bar{v}_f =(v_f^a+v_f^b)/2$ is the average Fermi velocity and we took, $\mu_{\alpha}=k_{F}^{2}/{2m_{\alpha}}$. For simplicity we considered a situation of perfect nesting for the unhybridized system, such that, the two bands cross the Fermi surface at the same $k_F$ for $V=0$ as shown in Fig.~\ref{fig1}.  The mismatch of the new Fermi wave-vectors of the hybridized bands is given by, $\delta k_{F}=k_{F}^{1 }-k_{F}^{2 }=2V/\bar{v}_{f}$. In the equation for the susceptibility we can neglect spin indexes since the normal system is paramagnetic.

In the normal system, at zero temperature, as the strength of the hybridization $V$ decreases, the condition $1-g \chi_V^{12}(q,0)=0$ in Eq.~\ref{gsinter} is eventually satisfied. In this case the interacting pair  susceptibility  $\tilde{\chi}(q,0)$ diverges implying that even in the absence of the fictitious field $(h_q=0)$ there is a spontaneous appearance of superconductivity in the system. The condition $1-g \chi_V^{12}(q,0)=0$ is a generalized Thouless criterion \cite{thouless} for a $q$-dependent superconducting instability. This instability can be triggered, as discussed below, either by decreasing the mismatch between the Fermi surfaces, i.e. reducing the hybridization, or by increasing the strength of the attractive interaction. The quantum phase transition from the normal to the $FF$ or helicoidal superconducting state at the critical value of hybridization $V_c$ (or $g_c$) is continuous and in the three dimensional case it's universality class has been obtained \cite{aline}. The value of the wave-vector $q_c$ characterizing the helicoidal superconducting state for $V <V_c(q)$ is that for which the condition  $1-g\chi_V^{12}(q,0)=0$ is first realized. It corresponds to that where $V_c(q)$ determined by the condition above has a maximum ($V_c(q_c)$ is a maximum). Due to the cylindrical symmetry of the problem any linear combination, $\Delta(\mathbf{r})=\sum_m \Delta_m e^{i \mathbf{q}_m \cdot \mathbf{r}}$ with $|\mathbf{q}_m|=q_c$ has the same critical value of hybridization \cite{shimahara}. This degeneracy for $V<V_c$ is removed by additional interactions or non-linear terms in the gap equation \cite{shimahara}. Any of these phases will have the same quantum critical behavior obtained below.

\subsection{The nature of the transition}
The non-interacting particle-particle, or pair dynamic susceptibility given by Eq.~\ref{chi2d} can be easily obtained at zero temperature.
It is given by
\begin{equation} \label{chi2dc}
\chi_V^{12}(q, \omega )=\rho \ln \left[\frac{\frac{2 \omega_c}{V}}{1- \frac{\omega }{2V} + \sqrt{(1-\frac{\omega }{2V})^2 -\bar{q}^2}}\right],
\end{equation}
where $\rho$ is the density of states at the Fermi level of the $2d$ system, $\bar{q}= \bar{v}_f q/2V$ and $\omega_c$ an energy cut-off.
The real part of the static susceptibility is given by,
\begin{eqnarray} \label{rechi2d}
{\rm Re}  \chi_V^{12}(q, \omega =0)&=&\rho \left[ \ln\frac{2 \omega_c}{h} - \ln[1+\sqrt{1-\bar{q}^2}] \right], \bar{q} \le 1 \nonumber \\
&=& \rho \left[ \ln\frac{2 \omega_c}{h} -\ln\bar{q} \right], \bar{q} >1.
\end{eqnarray}
The Thouless criterion for the appearance of superconductivity is given by, $1-g \Re e \chi_{V_c}^{12}(q,0)=0$ and coincides with the condition for the divergence of the static interacting pair susceptibility, $\tilde{\chi}(q,\omega =0)$ in Eq.~\ref{gsinter}.
\begin{figure}
\includegraphics[width=8cm]{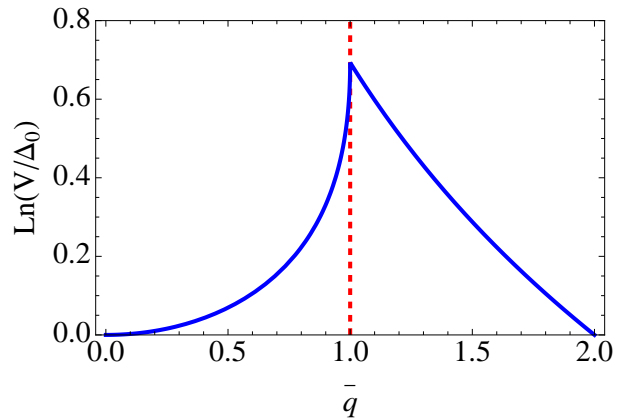}
\caption{(Color online) The logarithm of the function in Eq.~\ref{rechi2d}. This has a maximum for $\bar{q}=1$ with discontinuous derivatives at this point.}
\label{fig2}
\end{figure}

At zero temperature, for a fixed value of the interaction and momentum, as the hybridization $V$ is reduced, the normal system becomes unstable to a superconducting ground state at a critical value $V=V_c$ determined by the above condition. Using that the superconducting gap for $V=0$ can be written as, $\Delta_0=2 \omega_c e^{-1/g\rho}$, we obtain for $V_c(\bar{q})$,
\begin{eqnarray} \label{rechi2d}
V_c(\bar{q})&=&\frac{\Delta_0}{1+\sqrt{1-\bar{q}^2}}, \bar{q} \le 1 \nonumber \\
&=&\frac{\Delta_0}{\bar{q}}, \bar{q} >1.
\end{eqnarray}
The logarithm of this function is plotted in Fig.~ \ref{fig2}. It is continuous and has a a sharp maximum at $\bar{q}=1$. However, it's derivative is discontinuous and different from zero at the maximum.

\begin{figure}
\includegraphics[width=4.5cm]{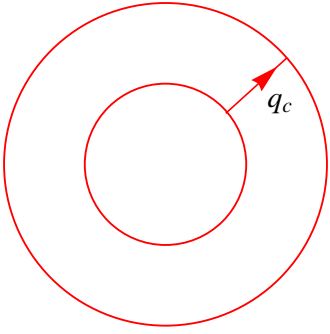}
\caption{(Color online) The wave-vector $q_c$ of the helicoidal superconductor is that joining the two Fermi surfaces at $k_F^1$ and $k_F^2$, i.e., $q_c=k_F^1-k_F^2$.}
\label{fig3}
\end{figure}

The value of $\bar{q}$ for which the instability first occurs is that for which $V_c(\bar{q})$ has a maximum, namely, $\bar{q}=\bar{q}_c=1$.
At this value of $\bar{q}$, $V_c =\Delta_0$. Also,  using that $k_{F}^{1 }-k_{F}^{2 }=2V/\bar{v}_{f}$, we get $q_c=k_{F}^{1 }-k_{F}^{2 }$, i.e., the wave vector of the instability is exactly that connecting the two Fermi surfaces at $k_F^1$ and $k_F^2$, as shown in Fig.~\ref{fig3}. This is also true in $1d$, but not in $3d$ \cite{aline}.

Next we expand the denominator of the interacting  dynamic pair susceptibility, Eq.~\ref{gsinter}, near $V=V_c$, $\bar{q}=\bar{q}_c$, and small frequencies $\omega $. For $q/\bar{q}_c>1$ and $1-2\omega /\bar{v}_f \bar{q}_c<\bar{q}/\bar{q}_c<1$, we get:
\begin{equation}
1-g  \chi_{V_c}^{12}(q,\omega )\approx g \rho \left[\frac{V-V_c}{V_c} + \frac{|\bar{q}-\bar{q}_c|}{\bar{q}_c} + i \sqrt{\frac{\omega }{V_c}}\right].
\end{equation}
This allows us to write the Gaussian part of the  action describing the normal-helicoidal superconductor quantum phase transition \cite{livro,mac}. It  is given by,
\begin{equation}\label{gaussian}
S\!=\!\!\int \! \! d\vec{q} \int \!\! d \omega  \! \left[ \frac{V\!-\!V_c}{V_c} + \frac{|q-q_c|}{q_c} +\sqrt{\frac{|\omega |}{V_c}} \right]\! |\Delta_q(q, \omega )|^2.
\end{equation}
The quantity $\delta=(V-V_c)/V_c$ measures the distance to the quantum critical point (QCP) associated with the normal-to helicoidal superconductor instability at $V_c= \Delta_0$ and with $q_c=2 \Delta_0/\bar{v}_f$. Since hybridization increases with pressure (P), near the QCP we can write $\delta=(P-P_c)/P_c$ where $P_c$ is the critical pressure below which the $q$-superconductor is stable. Within this Gaussian approximation, we can define a reduced correlation length $\xi/a=1/\delta$, where $a$ is the lattice spacing, such that, the zero temperature correlation length exponent takes the value $\nu=1$ for this model. Also, we can immediately identify  \cite{livro} that the dynamic critical exponent $z$ that scales the energy, or frequency, at the QCP is given by $z=2$. Even at the Gaussian level this quantum phase transition is in a new universality class since the critical exponents are non-standard Gaussian exponents. For example, we find $\nu=1$ instead of the usual Gaussian correlation length exponent $\nu_G=1/2$.

\begin{figure}
\includegraphics[width=8cm]{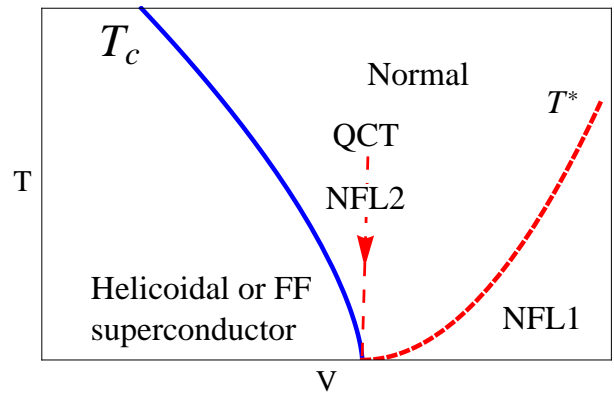}
\caption{(Color online) Phase diagram of the nearly two-dimensional metal near the normal-to-inhomogeneous superconductor quantum instability at $V_c$. The critical line $T_c \propto (V-V_c)^{2/3}$.   QCT (dashed line) refers to the quantum critical trajectory (see text). The  line $T_{*}$ marks the crossover between two different non-Fermi liquid behaviors, as discussed in the text.  The homogeneous superconductor part of the phase diagram for small $V$ and its boundary with the helicoidal or $FF$ phase are not shown in the figure since they are not studied here.}
\label{fig4}
\end{figure}

The free energy associated with the Gaussian action in two dimensions, Eq.~\ref{gaussian}, is given by \cite{livro},
\begin{equation} \label{free}
f = -\frac{2}{\pi} \frac{1}{(2 \pi)^2} T \int_0^{q_c} dq q  \int_0^{\infty} \frac{d \lambda}{e^{\lambda}-1} \tan^{-1}\left(\frac{\sqrt{\frac{\lambda T \xi^z}{V_c}}}{1+q \xi}\right).
\end{equation}
It can be written in the scaling form,
\begin{equation}
f \propto |\delta|^{\nu(d+z)}F[\frac{T}{T^*}],
\end{equation}
with $d=2$, $\nu=1$ and $z=2$. The characteristic temperature $T^*=V_c \xi^{-z}= V_c |\delta|^{\nu z}= V_c  |\delta|^{2}$.
In the non-critical side of the phase diagram in Fig.~\ref{fig4} there is a crossover temperature between two different regimes. For $T<< T^*$, the free energy is given by,
\begin{equation} \label{free}
f = \frac{-\zeta(3/2) }{4 \pi^{5/2}} \frac{T^{3/2}}{\sqrt{V_c}}   [q_c - \xi^{-1} \ln(1+q_c \xi)].
\end{equation} 
At  low temperatures,  $T \ll T^*$,  the specific heat $C/T = - \partial^2 f /\partial T^2$ has a power law behavior $C \propto \sqrt{T}$ with a coefficient that increases as the QCP is approached ($\xi^{-1} \propto \delta \rightarrow 0$). So at low temperatures before the instability, the normal system behaves as a non-Fermi liquid. For $T \gg T^{*}$, the specific heat beside the $C \propto \sqrt{T}$ term has  contribution $C/T \propto \ln T$. The latter is the scaling contribution $C/T \propto T^{(d-z)/z}$ which for $d=z$ gives rise to a logarithmic term.

Now let us assume that the action close to the QCP has an expansion in powers of the order parameter,
\begin{widetext}
\begin{eqnarray}
S&=&\int  d\vec{q} \int d \omega  \left[ \chi_G^{-1}(q,\omega )\right] |\Delta(\vec{q}, \omega )|^2  + \nonumber \\
&& \int \prod d\vec{q_i} \int \prod d \omega_{i} u(\{\vec{q}_i\},\{\omega_{i}\})\Delta(\vec{q_1},\omega_{1})\Delta(\vec{q_2},\omega_{2})\Delta(\vec{q_3},\omega_{3})\Delta(\vec{q_4},\omega_{4})\delta(\sum\vec{q}_i)\delta(\sum\omega_{i}),
\end{eqnarray}
\end{widetext}

\noindent with $i=1$ to $4$ and a quartic coefficient $u(\{\vec{q}_i\},\{\omega_{0i}\})$. If we neglect all $q$ and $\omega $ dependences of this term, such that, $u(\{\vec{q}_i\},\{\omega_{0i}\})=u_0$, simple power counting shows that 
the correlation function of the order parameter scales as $\Delta^{\prime 2}=b^{d+z-2+\eta}\Delta^2$ and  we find the critical exponent $\eta=1$.  Furthermore the quartic interaction among the fluctuations scales as:
\begin{equation}
u^{\prime}_0=b^{2 -(d+z)}u_0,
\end{equation}
such that these interactions are irrelevant, in the renormalization group sense, for $d+z \ge 2$. Then in the present case this implies that the Gaussian action gives the correct exponents describing the quantum critical point.  However  \cite{millis}, $u_0$ is dangerously irrelevant for $d+z >2$ implying departures from naive scaling \cite{livro}. For example, the shape of the temperature dependent critical line shown in Fig.~\ref{fig4} is given by, 
 $T_c(V) \propto (V_c-V)^{\psi}$ with the shift exponent, $\psi=\nu z/(1+\nu \theta_u)$, where $\theta_u=d+z-2$ is the scaling exponent of the dangerously irrelevant quartic interaction $u_0$ \cite{sachdev}. We obtain $\psi=z/(d+z-1)=2/3$ different from the mean-field result $\psi=1/2$ and from the naive scaling prediction $\psi=\nu z=2$ \cite{livro}. The correlation length along the quantum critical trajectory (QCT) in Fig.~\ref{fig4} diverges as $T \rightarrow 0$ as $\xi \propto (u_0 T^{1/\psi})^{-\nu}=u_{0}^{-1}T^{-3/2} $ \cite{bjp}, which makes manifest the dangerous irrelevant character of $u_0$.
This is different from the naive scaling result $\xi \propto T^{-1/z}$ which does not take into account the dangerous irrelevant nature of $u_0$.

We have given above a complete description of the normal-to-inhomogeneous superconductor quantum phase transition at the Gaussian level. As we have shown this is the correct theory of this quantum phase transition for a constant $u_0$ quartic interaction. However, the assumption of taking $u(\{\vec{q}_i\},\{\omega_{0i}\})=u_0$ may not be justified, particularly in two dimensions.  For the case the frequency-dependence of $u$ is discarded, but its momentum dependence is considered, it has been shown that the quartic interaction obtained from an expansion of the gap equation has a diverges at the wave-vector $\bar{q}_c=1$ of the superconducting instability \cite{shimahara,combescot}. At finite temperatures this divergence is removed and this also may occur for  finite frequencies or moderate disorder. A possible consequence of this anomalous behavior of $u$ is to change the order of the quantum phase transition  \cite{shimahara,combescot}.

Finally, it is interesting to point out that the correlation length exponent of the incipient FFLO phase ($\nu_{FFLO}=1$) being larger than that of any competing spin density wave ($\nu_{SDW}=1/2$) \cite{sachdev}, the former should override the latter instability.

\section{Discussion and Conclusion}

In multi-orbital metallic systems electrons from different bands coexist at the Fermi surface. These electrons have different Fermi wave-vectors and this difference provides a natural scale for inhomogeneous superconductivity in the case of inter-band attraction. The Fermi wave-vectors mismatch exists even in the absence of an external magnetic field and can be tuned by external pressure which directly changes the hybridization. Among the multi-orbital metals, the heavy fermion materials appear as strong candidates to exhibit the phenomenon investigated here. The dominant interaction in these compounds is the Kondo coupling between itinerant and localized electrons and this naturally leads to an attractive inter-band interaction. Many of these systems have tetragonal structures with cylindrical Fermi surfaces for which the results we have obtained directly apply. Also, two-dimensions favors the inhomogeneous FFLO type of phases \cite{fflo2d} and in addition the mechanism studied here avoids the complications due to orbital effects since it does not require an external magnetic field. Heavy fermions are in general very sensitive to external pressure and this allows to probe a large region in their phase diagrams. For $Ce$ based heavy fermion systems, in many cases, at least two superconducting phases appear as they are driven with increasing pressure from antiferromagnetic, superconducting and normal states \cite{jaccard}. The $FF$ or helicoidal superconducting state is a candidate for the first superconducting instability as pressure is reduced from the normal state. This transition occurs at a superconducting quantum critical point (SQCP) whose properties and universality class we have obtained. 
The theoretical description of this quantum phase transition requires considering  time and space as fundamental dimensions \cite{livro}. This is implied by the uncertainty principle which must remain scale invariant leading to the inextricability of these quantities.  We have characterized the thermodynamic behavior near the QCP and this should be useful for identifying the helicoidal phase from its precursor effects in actual systems.

The peculiar quantum critical behavior of our model is a consequence that $\bar{q}_c=1$ is a singular point of the pair susceptibility.
In many real systems, we can observe consequences of the structure of Lindhard functions particularly in low dimensional systems, as the Kohn anomaly and the Peierls instability in one-dimensional conductors \cite{smith}. It is reasonable than to expect that our results should hold for real materials.

Finally, we point out that we have not explored the phase diagram for small mismatches where a homogeneous BCS phase exists up to a critical mismatch. This problem has been treated in Refs. \cite{fflo2d,Simmons,Rainer,combescot}.

{\it Note added in proof.} Recently, we became aware of the paper~\cite{Samokhin} where they calculate the susceptibility close to the quantum normal-FFLO transition.

\section*{ACKNOWLEDGMENTS}
H.C. and M A.C. are partially supported by CNPq. The authors also acknowledge partial support from FAPEMIG and FAPERJ, respectively.
M.A.C. would like to thank Martin Garst for useful discussions. H.C. acknowledges the kind hospitality at CBPF where this work was done.

\end{document}